\documentclass[graybox,natbib,nosecnum]{svmult}
\bibpunct{(}{)}{;}{a}{}{,} % suppress commas between author-names and year

\pdfoutput=1   %forces use of pdflatex. Disable if you prefer to use .eps or .ps figures.
\usepackage{mathptmx}       % selects Times Roman as basic font
\usepackage{helvet}         % selects Helvetica as sans-serif font
\usepackage{courier}        % selects Courier as typewriter font
\usepackage{type1cm}        % activate if the above 3 fonts are
\usepackage{makeidx}         % allows index generation
\usepackage{graphicx}        % standard LaTeX graphics tool
\usepackage{multicol}        % used for the two-column index
\usepackage[bottom]{footmisc}% places footnotes at page bottom
\usepackage[normalem]{ulem}	% for strike-through of text with \sout{}  
\usepackage{hyperref}  %for hyperlinks

\usepackage{soul}   % for high-lighting of text
\newcommand*\aap{A\&A}

\newcommand*\aapr{A\&A Rev}

\newcommand*\aj{AJ}

\newcommand*\apj{ApJ}
\newcommand*\apjl{ApJ}

\newcommand*\araa{ARA\&A}

\newcommand*\grl{Geophys Res Lett}

\newcommand*\icarus{Icarus}

\newcommand*\mnras{MNRAS}

\newcommand*\nat{Nature}

\newcommand*\pasj{PASJ}

\newcommand*\solphys{Sol Phys}

\newcommand{\hbindex}[1]{\index{#1}}  %highlights index entries

\usepackage{amssymb}
\newcommand{\msano}{{\rm M}_\odot ~{\rm yr}^{-1}}
\newcommand{\mdot}{\dot{M}}

\makeindex             % used for the subject index
\begin{document}

\title*{Stellar Coronal and Wind Models: Impact on Exoplanets}
% Use \titlerunning{Short Title} for an abbreviated version of
% your contribution title if the original one is too long
\author{Aline A.~Vidotto}
% Use \authorrunning{Short Title} for an abbreviated version of
% your contribution title if the original one is too long
\institute{School of Physics, Trinity College Dublin, the University of Dublin, Dublin-2, Ireland, \email{aline.vidotto@tcd.ie}}
%
% Use the package "url.sty" to avoid
% problems with special characters
% used in your e-mail or web address
%
\maketitle

\abstract{
Surface magnetism is believed to be the main driver of coronal heating and stellar wind acceleration. Coronae are believed to be formed by plasma confined in closed magnetic coronal loops of the stars, with winds mainly originating in open magnetic field line regions. In this Chapter,  we review some basic properties of stellar coronae and winds and present some existing models. In the last part of this Chapter, we discuss the effects of coronal winds on exoplanets.}

%%%%%%%%%%%%%%%%%%%%%%%%%%%%%%%%%%%%%%%%%%
\section{Introduction}
About $90\%$ of the currently known exoplanets orbit around low-mass stars. These stars  ($0.1 \lesssim M_\star/M_\odot \lesssim 1.3$), while in the main-sequence phase, have convective interiors that vary in extension as a function of the stellar mass. Below $\sim 0.4 M_\odot$, these stars are fully convective. Above this mass threshold, there is an appearance of a radiative core, whose size is larger for more massive stars. In turn,  the convective part of the star is limited to the outer layers and becomes progressively smaller as one goes towards more massive  stars. At $\sim 1.3 M_\odot$, the outer convective envelope is already very small.

As convection is one of the key ingredients in the generation of magnetic fields, main-sequence low-mass stars have \hbindex{surface magnetic fields}. This magnetism gives rise to a multitude of phenomena, from small and localised features (spots, active regions, prominences) to large-scale ones (global magnetism, coronal holes, helmet streamers).

Surface magnetism is also believed to be the main driver of coronal heating and stellar wind acceleration. However, at present, there is no consensus of the basic physical mechanisms involved in these processes. Even for the Sun, heating of the solar corona and acceleration of the solar wind are still currently being debated, with possible scenarios relating to propagation and dissipation of waves and turbulence in open  magnetic flux tubes and/or reconnection between open and closed magnetic flux tubes \citep{2009LRSP....6....3C}.

In this Chapter, we start by reviewing basic properties of stellar coronae and winds. We then present a review of some existing models. The last part of this Chapter is dedicated to the impact of coronal winds on exoplanets.

%%%%%%%%%%%%%%%%%%%%%%%%%%%%%%%%%%%%%%%%%%
\section{Observationally-derived properties of stellar coronae}
Low-mass stars harbour hot coronae with average temperatures on the order of $10^6$ -- $10^7$~K \citep{2004A&ARv..12...71G,2005ApJ...622..653T,2015A&A...578A.129J}. The hot stellar coronae are detected in X-ray wavelengths \citep[e.g.][]{2003A&A...397..147P,2004A&ARv..12...71G,2005ApJ...622..653T,2011A&A...527A.144M,2011ApJ...743...48W,2013A&A...552A...7S,2014ApJ...785..145P,2015A&A...578A.129J}, during both quiescent and flaring states. Coronae are believed to be formed by plasma confined in closed magnetic coronal loops of the stars. An indication that coronae have indeed their origins in stellar magnetism comes from the observed correlation between X-ray emission and stellar magnetic fields \citep{2003ApJ...598.1387P, 2014MNRAS.441.2361V}. In this Section, we highlight a few observed properties of stellar coronae. An interested reader will find comprehensive reviews of  X-ray \hbindex{stellar coronae}  in, e.g.,  \citet{2004A&ARv..12...71G,2009A&ARv..17..309G,2015RSPTA.37340259T}.

\runinhead{X-ray coronae and stellar rotation:} Earlier studies have shown the connection between \hbindex{stellar rotation} and chromospheric activity \citep{1967ApJ...150..551K}. Similarly, X-ray emission has  also been recognised to correlate with stellar rotation, with the exception of fast-rotating stars \citep[e.g.][]{1981ApJ...248..279P,2003A&A...397..147P,2011MNRAS.411.2099J,2011ApJ...743...48W,2014ApJ...794..144R}. For this reason, the activity-rotation relation is usually divided into two parts. Fast-rotating stars have X-ray emission that is roughly independent of rotation. They are in the so called {\it saturated regime}. These stars have X-ray luminosities that account for about $0.1\%$ of their bolometric luminosities. For slower rotators, in the {\it unsaturated regime}, X-ray luminosities increases with rotation rate $\Omega_\star$ as \citep{2014ApJ...794..144R}
\begin{equation}\label{eq.lx}
L_x \propto \Omega_\star^{2.01 \pm 0.05} .
\end{equation}
The  rotation rate at which stars transition from unsaturated to saturated regimes corresponds to \citep{2015A&A...578A.129J}
\begin{equation}
\frac{\Omega_{\star, {\rm sat}}}{\Omega_\odot} \simeq 13.53 \left( \frac{M_\star}{M_\odot} \right)^{1.08}, 
\end{equation}
where $\Omega_\odot = 2.67 \times 10^{-6}$ rad~s$^{-1}$. The saturation threshold is mass-dependent, with lower-mass stars transitioning from the saturated to the unsaturated regime at lower rotation rates. As rotation decreases with the square-root of the age of stars \citep{1972ApJ...171..565S}, stars in the lowest-mass range (e.g., M dwarfs) remain saturated even at relatively old ages (note also that these stars have longer lifetimes).

Another observed link between rotation and X-ray emission is seen in X-ray lightcurves. Because  X-ray emission arises in closed magnetic coronal loops and since the distribution of closed/open magnetic field line regions at the surface of the star is inhomogeneous, stars can also show rotational modulations in X-ray \citep{2005ApJ...621..999H,2007MNRAS.377.1488H}.

\runinhead{X-ray coronae and temperatures} It has also been shown that stars with hot coronae have high X-ray emission \citep[e.g.][]{2005ApJ...622..653T}. For low-mass main-sequence stars, there is a tight relation between X-ray flux $F_x$  and average \hbindex{coronal temperature} $\tilde{T_c}$ \citep{2015A&A...578A.129J}
\begin{equation}\label{eq.Tc}
F_x= 0.9 \left( \frac{\tilde{T_c}}{10^6~{\rm K}} \right)^{3.8} {\rm erg~cm}^{-2}{\rm ~s}^{-1}.
\end{equation}
This empirical relation is useful for estimating the average coronal temperature of stars, once $F_x$ is known. $F_x$ can either be determined observationally or by using the rotation-activity relation (e.g., Equation \ref{eq.lx}). As we will see in this Chapter, the \hbindex{stellar wind} temperature is an unknown in the models. Models that relate the temperature of the wind to the temperature of the corona can benefit from the empirical relation (\ref{eq.Tc}).

\runinhead{X-ray coronae and magnetism} The link between coronae and magnetism has long been identified. For this reason, X-ray emission is often used as a proxy for stellar magnetism. One way to validate this is by confronting observed values of X-ray luminosities/fluxes with observations of stellar magnetism. 

Two methods are mostly used to measure stellar magnetism. The Zeeman-induced line broadening of unpolarised light (Stokes I), or Zeeman broadening (ZB) technique \citep[e.g.,][]{1994ASPC...64..477S,1996IAUS..176..237S,1999ApJ...510L..41J,2007ApJ...664..975J,2001ASPC..223..292S,2009ApJ...692..538R}, yields estimates of the average of the total unsigned surface field strength  $\langle |B_I| \rangle$ (small- and large-scale structures). This technique does not provide information of the topology of the field. The \hbindex{Zeeman Doppler imaging} (\hbindex{ZDI}) technique (Stokes V), on the other hand, is able to reconstruct the intensity and topology of the stellar magnetic field \citep[e.g.,][]{1997A&A...326.1135D,2009ARA&A..47..333D,2013AN....334...48M}, but cannot reconstruct the small-scale field component, which is missed within the resolution element of the reconstructed ZDI maps \citep{2010MNRAS.404..101J,2011MNRAS.410.2472A,2014MNRAS.439.2122L}. As a consequence, the ZDI magnetic maps are limited to measuring large-scale magnetic field. 

\begin{figure}
\includegraphics[width=5.9cm]{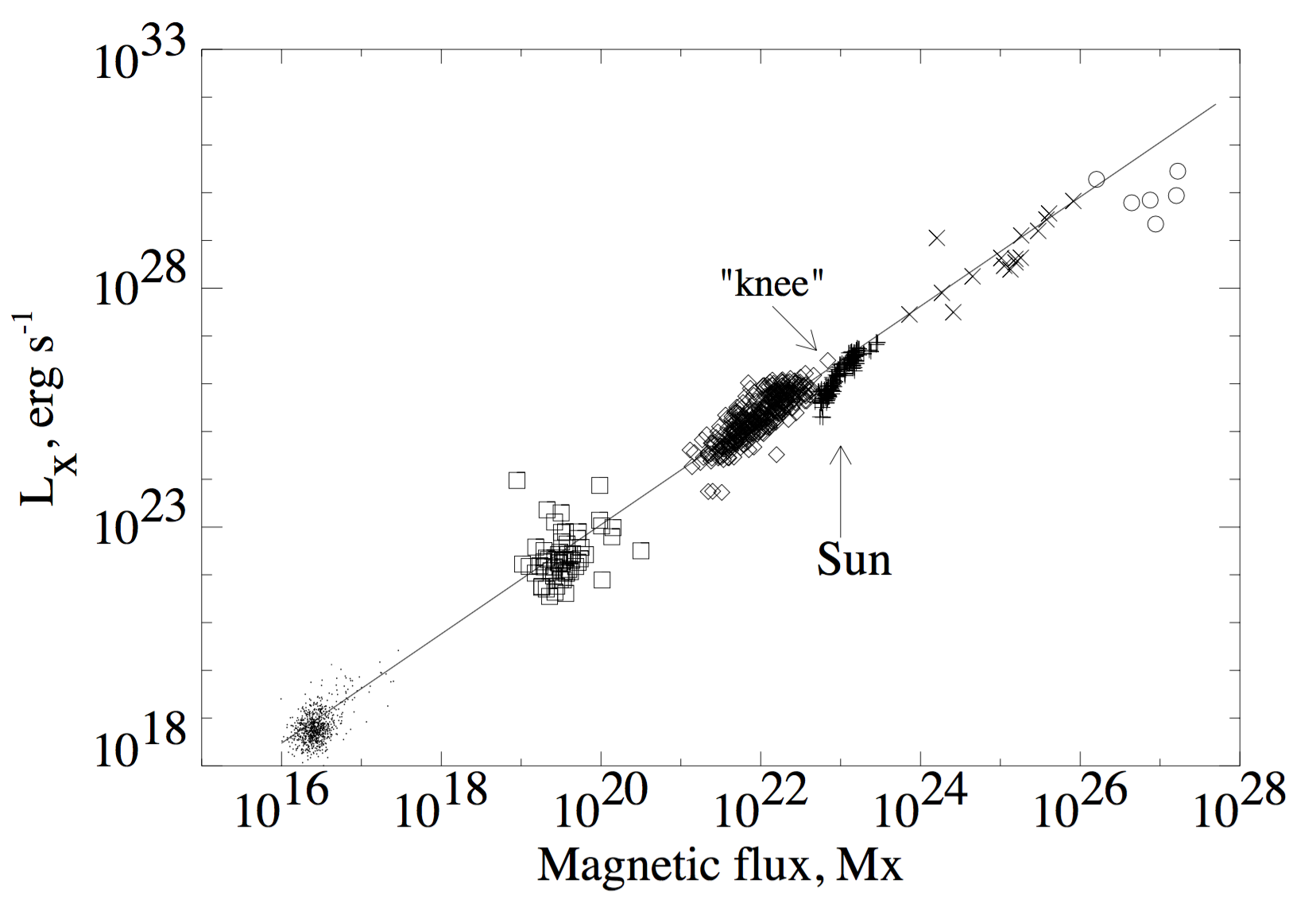}
\includegraphics[width=5.5cm]{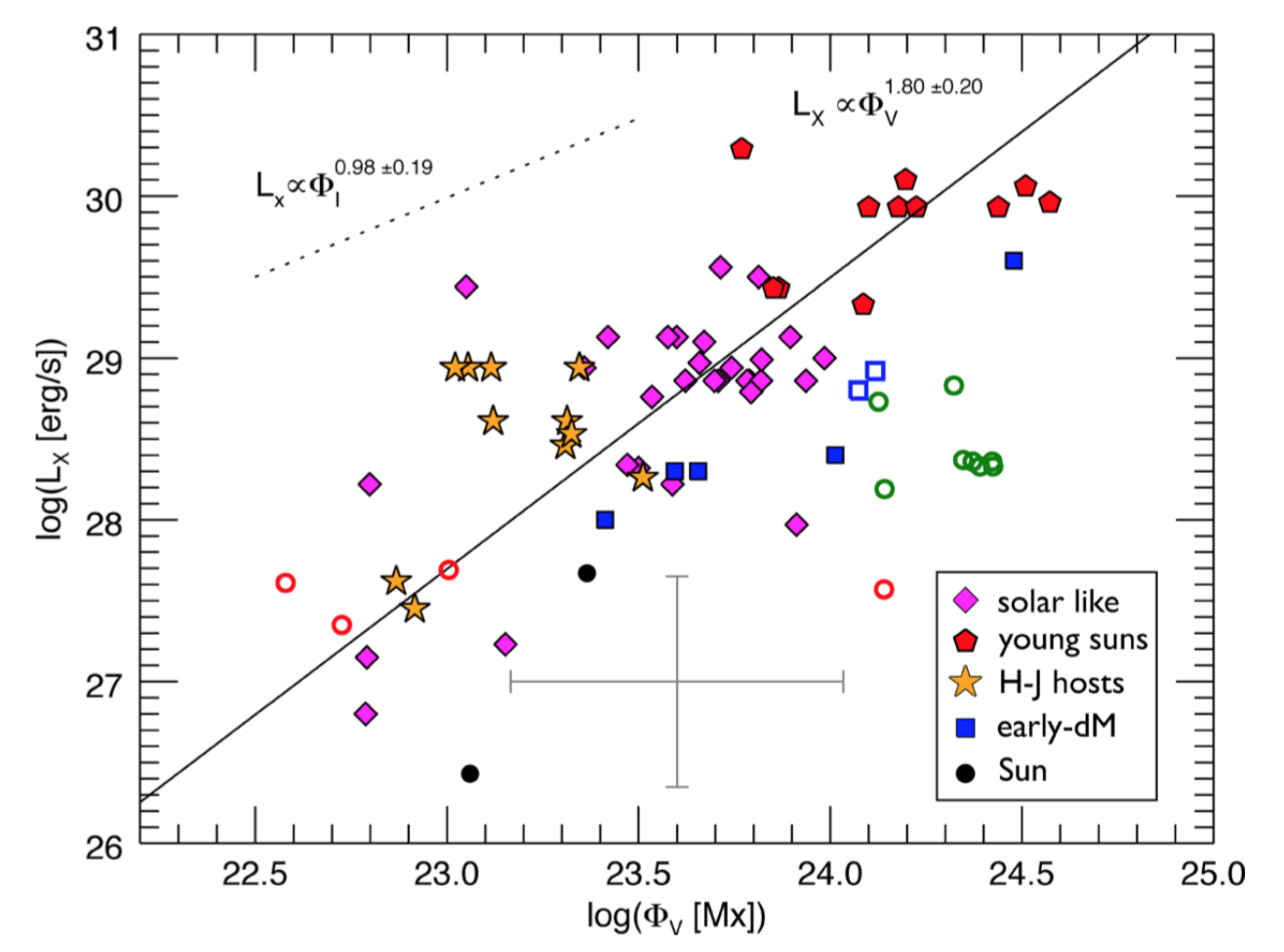}
\caption{(a) Relation between the X-ray luminosities and the unsigned magnetic fluxes $\Phi_I$ measured by the Zeeman broadening technique \citep{2003ApJ...598.1387P}. Reproduced by permission of the AAS.  
(b) The same as in (a) but for magnetic field measurements done with the Zeeman Doppler imaging technique. The different symbols mainly indicate different types of surveys. The solid line shows the empirical fit through the data, while the dashed line (at an arbitrary vertical offset) is indicative of the slope found by the Zeeman broadening technique, when considering the 16 G, K, M stars. From Vidotto et al. (2014 MNRAS, 441, 2361).}
\label{fig:av_Lx_magfield}       
\end{figure}

\citet{2003ApJ...598.1387P} found that the X-ray luminosities are related to the unsigned magnetic fluxes $\Phi_I$ measured by the ZB technique
\begin{equation}
L_x \propto \Phi_I^{1.13\pm 0.05}, 
\end{equation}
where 
$$\Phi_I = \langle |B_I| \rangle  4 \pi R_\star^2 . $$
To derive this relation, \citet{2003ApJ...598.1387P} considered magnetic field observations of the Sun (X-ray bright points, active regions, quiet Sun and integrated solar disk) and pre- and main-sequence stars. This empirical relation can be seen in Figure \ref{fig:av_Lx_magfield}a, spanning about $12$ orders of magnitude in magnetic flux. 

Similarly, \citet{2014MNRAS.441.2361V} found that
\begin{equation}
L_x \propto \Phi_V^{0.913 \pm 0.054}, 
\end{equation}
where 
$$\Phi_V = \langle |B_V| \rangle  4 \pi R_\star^2  $$
is the unsigned magnetic flux as derived from the \hbindex{ZDI} technique (i.e., only contains the large-scale component of the stellar magnetic field). To be consistent with the method from \citet{2003ApJ...598.1387P}, the relation above considers both main-sequence stars and pre-main sequence (accreting) stars. The slope found by \citet{2014MNRAS.441.2361V} is  consistent to the nearly linear trend found by \citet{2003ApJ...598.1387P}.  

Figure \ref{fig:av_Lx_magfield}a\footnote{The data provided in Figure \ref{fig:av_Lx_magfield}a are from: \cite{1999MNRAS.302..437D,2003MNRAS.345.1145D,2008MNRAS.390..545D,2008MNRAS.385.1179D,2008MNRAS.386.1234D,2010MNRAS.409.1347D,2010MNRAS.402.1426D,2011MNRAS.412.2454D,2011MNRAS.417..472D,2011MNRAS.417.1747D,2012MNRAS.425.2948D,2013MNRAS.436..881D, 2006MNRAS.370..468M,2011MNRAS.413.1922M,2007MNRAS.374L..42C,2008MNRAS.384...77M,2008MNRAS.390..567M,2010MNRAS.407.2269M,2008MNRAS.388...80P,2009A&A...508L...9P,2009MNRAS.398..189H,2009MNRAS.398.1383F,2010MNRAS.406..409F,2012MNRAS.423.1006F,2013MNRAS.435.1451F,2011AN....332..866M,2012A&A...540A.138M,2011MNRAS.413.1949W, 2015MNRAS.449....8W,2017MNRAS.465.2076W, 2016ApJ...820L..15D, 2016MNRAS.457..580F} and from Petit et al. in prep.} shows that the pre-main sequence stars (open circles) are under-luminous as compared to the empirical fit (solid line). When considering only the sample of $16$ G, K and M dwarf stars (i.e., no solar data nor accreting PMS stars), \citet{2003ApJ...598.1387P} found that $L_x^{\rm (dwarfs)} \propto \Phi_I^{0.98\pm 0.19}$. Considering the same types of objects, the relation derived from ZDI data yields $L_X^{\rm (dwarfs)} \propto \Phi_V^{1.80\pm 0.20}$ (based on a larger sample of {$61$} dwarf stars). This is shown in Figure \ref{fig:av_Lx_magfield}b. Given the larger errors in the exponents of the fits, both relations are consistent to each other within $3\sigma$. Still, this is a topic worth of future investigation. For example, finding a different power law for $\Phi_V$ and $\Phi_I$ might clarify on how the small-scale and large-scale field structures contribute to X-ray emission.

%%%%%%%%%%%%%%%%%%%%%%%%%%%%%%%%%%%%%%%%%%
\section{Observationally-derived properties of stellar winds}
Low-mass stars undergo mass loss through winds during their entire lives. Contrary to  the Sun, in which the solar wind can be probed in situ, the existence of winds around low-mass stars is known indirectly, e.g., from the observed rotational evolution of stars \citep[e.g.,][]{2014prpl.conf..433B}. Measuring the wind \hbindex{mass-loss rates} $\dot{M}$ of  cool, low-mass stars is challenging, as these winds are rarefied and difficult to be directly detected. 

Attempts to measure low mass star's winds have been done through radio observations of their free-free thermal emission at radio wavelengths \citep[e.g.,][]{1997A&A...319..578V,2000GeoRL..27..501G,2014ApJ...788..112V} and through X-ray observations of the emission generated when ionised wind particles exchange charges with neutral atoms of the interstellar medium \citep{2002ApJ...578..503W}. Other attempts involve the observations of coronal radio flares \citep{1996ApJ...462L..91L} or the accretion of wind material from a cool low-mass star to a white dwarf in binary systems \citep{2006ApJ...652..636D,2012MNRAS.420.3281P}. So far, the indirect method proposed by \citet{2001ApJ...547L..49W}, which involves reconstruction of stellar Lyman-$\alpha$ line (see below), has been the most successful one, enabling estimates of $\dot{M}$ for about a dozen dwarf stars. 
To illustrate the challenging aspects of measuring \hbindex{mass-loss rates}, we show in Table~\ref{table.proxima_centauri}  tentative measurements of $\mdot$ of the closest star to us, namely Proxima Centauri. Recently, the interest in understanding Proxima Centauri has increased due to the discovery of a terrestrial type planet orbiting in its habitable zone \citep{2016Natur.536..437A}.

\begin{table}
\caption{Characteristics of Proxima Centauri and its wind.}
\label{table.proxima_centauri}      
\begin{tabular}{p{4cm}p{2.4cm}p{2cm}}
\hline\noalign{\smallskip}
Physical property  & Value & Reference  \\
\noalign{\smallskip}\svhline\noalign{\smallskip}
mass ($M_\odot$) & $0.123$  & a \\
radius ($R_\odot$) & $0.141$  & b,c \\
rotation period (days) & $\sim 83$  & c \\
$F_x$ ($10^6 $erg~cm$^{-2}$s$^{-1}$)& $\sim 1.2$  & d \\
$\tilde{T_c} (10^6~{\rm K})$ & $2.7$ & e\\
spectral type & M5.5 & f \\
total magnetic flux (G) & $600$ & g \\
$\mdot (\mdot_\odot = 2\times 10^{-14}~ \msano)$ & $<350$ & h \\
\ldots & $<14$ & i \\
 \ldots & $0.2$ & f \\
\noalign{\smallskip}\hline\noalign{\smallskip}
\end{tabular}\\
a: \cite{2016A&A...596A.111R}; b: \cite{2009A&A...505..205D}; c: \cite{2016Natur.536..437A}; d: \cite{2004LRSP....1....2W}; e: \cite{2004A&A...416..713G}; f: \cite{2001ApJ...547L..49W}; g: \cite{2008A&A...489L..45R}; h:  \cite{1996ApJ...460..976L}; i: \cite{2002ApJ...578..503W}. 
\end{table}

\begin{figure}
\includegraphics[width=11.5cm]{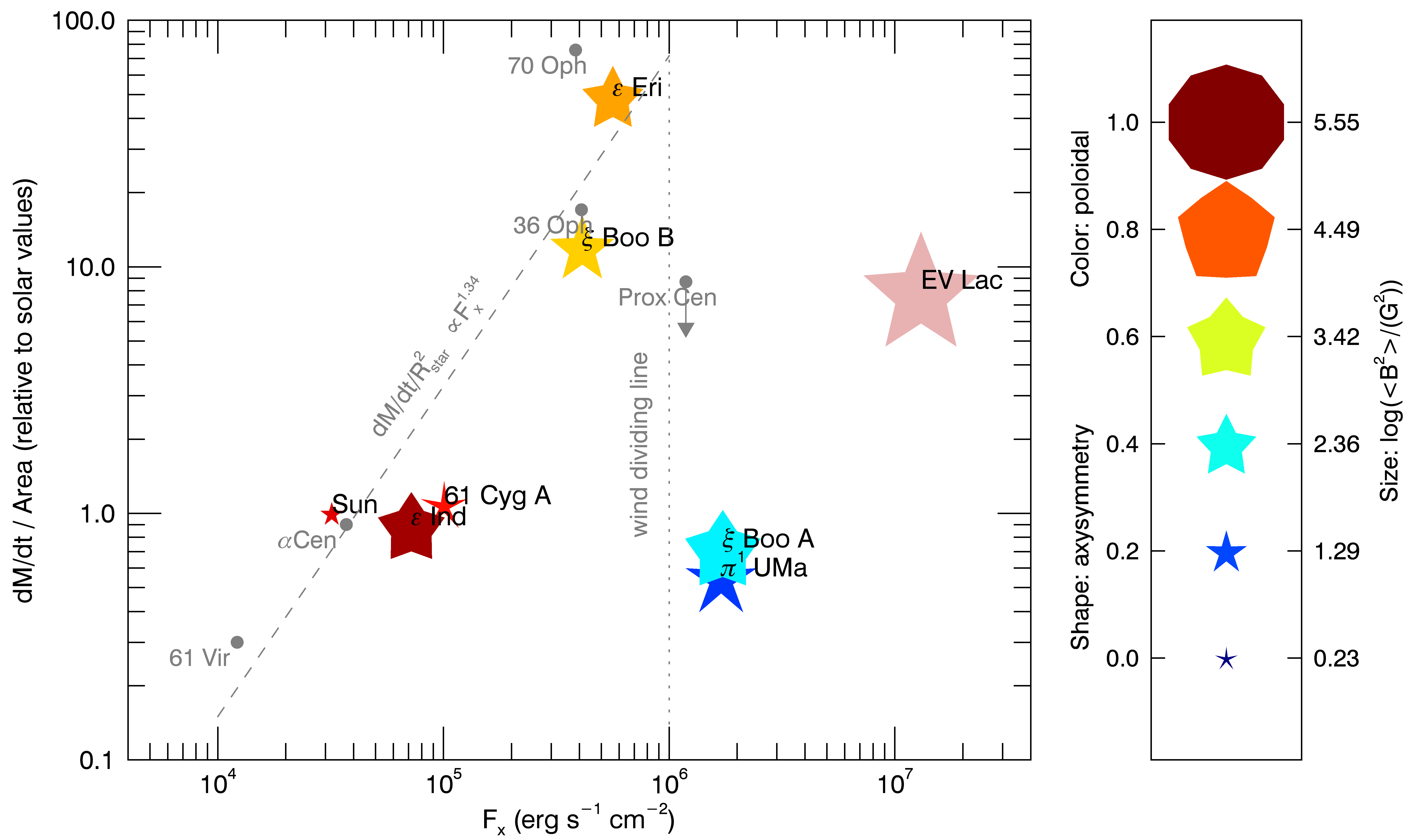}
\caption{Compilation of the mass-loss rates derived by \citet{2004LRSP....1....2W,2014ApJ...781L..33W} as a function of \hbindex{X-ray fluxes}. Grey circles are stars that do not yet have measurements of their surface magnetic fields with ZDI. For the other systems,  symbol sizes are proportional to the magnetic energy $\log \langle B^2 \rangle$, their colours indicate the fractional poloidal energy (ranging from deep red for purely poloidal field $f_{\rm pol}=1$ to blue for purely toroidal field $f_{\rm pol}=0$), and their shapes indicate the fraction of axisymmetry of the poloidal component (ranging from a decagon for purely axisymmetric field $f_{\rm axi}=1$ to a point-shaped star for $f_{\rm axi}=0$). The $\mdot$ -- $F_x$ relation (Eq.~\ref{eq.wood}) is shown as a dashed line and the WDL at $F_x=10^6$~erg~cm$^{-2}$s$^{-1}$ is shown as a dotted line. The \hbindex{ZDI} maps used to produce this figure are from \cite{2008MNRAS.390..567M, 2012A&A...540A.138M, 2014A&A...569A..79J, 2016A&A...594A..29B, 2016MNRAS.459.1533V}; Petit et al in prep; and Boisse et al in prep. Figure based on Vidotto et al. (2016, MNRAS, 455L, 52).}
\label{fig:av_astrosphere}       % Give a unique label
\end{figure}

Wood et al.'s method explains the  excess absorption observed in the blue wing of the Lyman-$\alpha$ line as caused by the hydrogen wall that forms during the interaction between the stellar wind and the interstellar medium. \citet{2002ApJ...574..412W,2005ApJ...628L.143W} found a relation between the mass-loss rate per unit surface area $A_\star=4\pi R_\star^2$ and the \hbindex{X-ray flux} for low-mass stars
\begin{equation}\label{eq.wood}
\dot{M}/4\pi R_\star^2 \propto F_x^{1.34}.
\end{equation}
As X-ray emission is related to rotation and rotation can be related to stellar ages, Equation (\ref{eq.wood}) implies that younger stars have higher $\dot{M}$ than older ones. Also, Equation (\ref{eq.wood}) is only valid for $F_x \lesssim 10^6$~erg~cm$^{-2}$s$^{-1}$ or ages $\gtrsim 600$~Myr. Figure \ref{fig:av_astrosphere} compiles the mass-loss rates derived by \citet{2004LRSP....1....2W,2014ApJ...781L..33W}. 

The break in the $\mdot$ -- $F_x$ relation found by \cite{2005ApJ...628L.143W} for active stars with $F_x\gtrsim 10^6$~erg~cm$^{-2}$s$^{-1}$ has been suggested to be caused by the topology of  \hbindex{stellar magnetic fields} that would inhibit the wind generation \citep{2005ApJ...628L.143W,2010ApJ...717.1279W}. \citet{2016MNRAS.455L..52V} analysed this hypothesis with a sub-sample of stars observed by Wood et al that also had observationally-derived large-scale magnetic fields with ZDI. These authors did not find any particular evidence that the magnetic field characteristics show an abrupt change at the wind dividing line (WDL, at $F_x \sim 10^6$~erg~cm$^{-2}$s$^{-1}$). In general, solar-type stars to the right of the WDL (namely $\xi$ Boo A and $\pi^1$ UMa) have higher fractional toroidal fields (blueish points in Fig.~\ref{fig:av_astrosphere}), but no break or sharp transition was found.

Very active stars, in particular, show variability in their \hbindex{magnetic field} properties on timescales on the order of a few years and can, for example, jump between states with highly toroidal fields and mostly poloidal fields (\citealt{2009A&A...508L...9P, 2012A&A...540A.138M, 2014A&A...569A..79J, 2015A&A...573A..17B}). If magnetic fields are to affect stellar winds,  significant scatter in the points to the right of the WDL are to be expected.  

%%%%%%%%%%%%%%%%%%%%%%%%%%%%%%%%%%%%%%%%%%
\section{Models of stellar coronal winds}\label{sec.av_models}
\begin{figure}
\includegraphics[width=8cm]{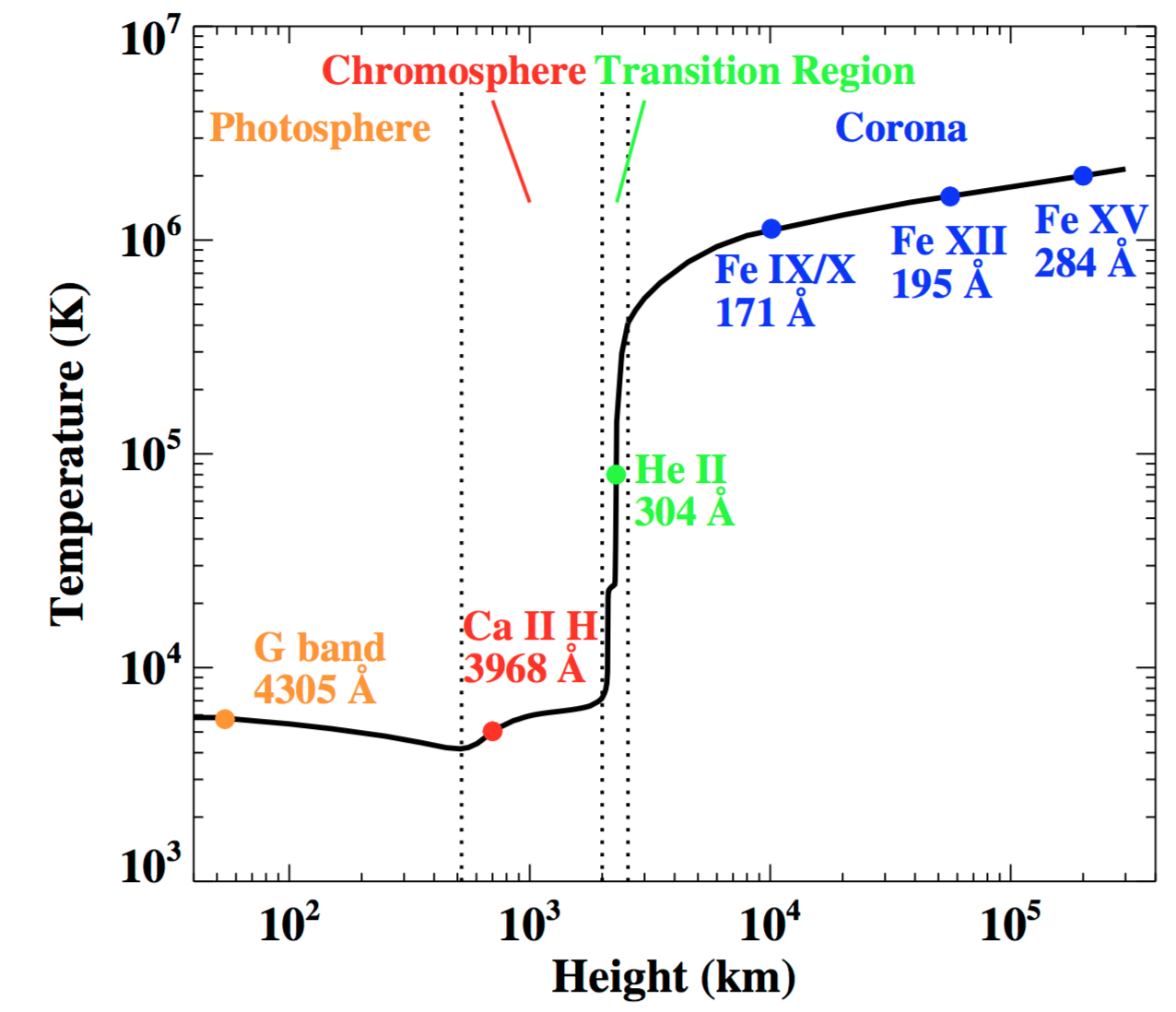}
\caption{Temperature variation with height for the solar atmosphere. Credit: Yang et al., A\&A, 501, 745, 2009, reproduced with permission \copyright ESO.
}
\label{fig.av_corona}       
\end{figure}

Studies of the solar wind have provided insights into the winds of low-mass stars. However, as there is still no consensus of the basic physical mechanisms involved in the acceleration of the solar wind \citep{2009LRSP....6....3C}, this uncertainty also propagates to models of stellar winds. The \hbindex{winds of low-mass stars} are believed to be magnetically-driven, in which coupling between stellar magnetism and convection transports free magnetic energy, which in turn is converted into thermal energy in the upper atmosphere of stars \citep{2014MNRAS.440..971M}, giving rise to a hot corona (illustrated in Figure \ref{fig.av_corona} for the solar atmosphere). The scale height of X-ray emitting stellar corona is likely to vary with the properties of the star \citep{2004A&A...414L...5J,2004A&ARv..12...71G}. A possible scenario to convert magnetic  into thermal energy involves the dissipation of waves and turbulence \citep[e.g.,][]{1983ApJ...275..808H, 2008ApJ...689..316C, 2011ApJ...741...54C, 2013PASJ...65...98S, 2014MNRAS.440..971M}. In addition to depositing energy,  waves also transfer momentum to the wind, accelerating it \citep[e.g.,][]{2006ApJ...639..416V}. In general, two modelling approaches are used in the study of the hot coronal winds of low-mass stars. We describe them next.

\runinhead{Self-consistent heating/acceleration mechanism}
The first approach involves a more rigorous computation of the wave energy and momentum transfer, i.e., the computations are done from ``first principles'' \citep[e.g.,][]{1973ApJ...181..547H, 1983ApJ...275..808H, 1980ApJ...242..260H, 1989A&A...209..327J, 2006ApJ...639..416V, 2006MNRAS.368.1145F, 2008ApJ...689..316C, 2011ApJ...741...54C, 2013PASJ...65...98S}. In these models, the increase in temperature from the colder photosphere to the hotter corona arises naturally in the solution of the equations as does the wind acceleration. Most of the models that treat the \hbindex{stellar wind} acceleration starting from the photosphere have been limited to analytical, one- and two-dimensional solutions, as, depending on the level of details of the physics involved in the wind acceleration/heating mechanism, models  can become computationally intensive. In particular, a challenging numerical aspect is the large density contrast between the photosphere and the rarefied corona  \citep[e.g.][]{2012ApJ...749....8M}. Additionally, models are usually restricted to the inner-most part of the wind and they usually adopt simple topologies for the stellar magnetic field. 

\runinhead{Global wind models} 
The second approach adopts a simplified energy equation, usually assuming the wind is described by a polytropic equation of state. In the latter, the thermal pressure $p_{\rm th}$ is related to density $n$ as  $p_{\rm th} \propto n^\gamma$, where $\gamma$ is known as the \hbindex{polytropic index}.  In these models, the computation often starts at the point where the temperature has already reached coronal values $\sim 10^6$~K \cite[e.g.,][]{1968MNRAS.138..359M, 1971SoPh...18..258P,1989ApJ...342.1028T, 1993MNRAS.262..936W, 2000ApJ...530.1036K,2001A&A...371..240L, 2005ApJ...632L.135M,2012ApJ...754L..26M,2009ApJ...699..441V,2009ApJ...703.1734V,2010ApJ...720.1262V,2009ApJ...699.1501C,2010ApJ...721...80C,2011ApJ...737...72P, 2015A&A...577A..28J,2015A&A...577A..27J,2015ApJ...798..116R, 2015ASSL..411...37L}. This approach ignores the physical reason of what led temperatures to increase from photospheric to coronal values. On the other hand, equations are simpler, allowing us to perform three-dimensional numerical simulations of ``global'' stellar winds, i.e., extending out to large distances from the star \citep{2009ApJ...699..441V,2009ApJ...703.1734V,2010ApJ...720.1262V,2012MNRAS.423.3285V, 2014MNRAS.438.1162V,2015MNRAS.449.4117V,2009ApJ...699.1501C,2010ApJ...721...80C,2013MNRAS.431..528J,2013MNRAS.436.2179L,2015ApJ...815..111S,2016MNRAS.459.1907N}. 

More recently, there have been efforts in developing a hybrid approach that combines the two approaches described above to study \hbindex{winds of low-mass stars}. In these hybrid models, a phenomenological approach of the (solar-based) wave heating mechanism is implemented in three-dimensional simulations of solar/stellar winds, starting from the upper chromosphere \citep{2010ApJ...725.1373V,2014ApJ...782...81V, 2013ApJ...764...23S,2015ApJ...813...40G, 2016A&A...588A..28A}. These models present a step forward in the modelling of winds of low-mass stars, as they, for example, do not need to impose a polytropic index to mimic energy deposition in the wind. However, there are still  some parameters that need to be imposed {\it a priori}, such as the energy flux of waves at the inner boundary and its dissipation length scale, as discussed in \citet{2013ApJ...764...23S}. In the case of the solar wind, these free parameters can be constrained from observations \citep{2013ApJ...764...23S,2014ApJ...782...81V}.

\begin{figure}
\includegraphics[width=5.8cm]{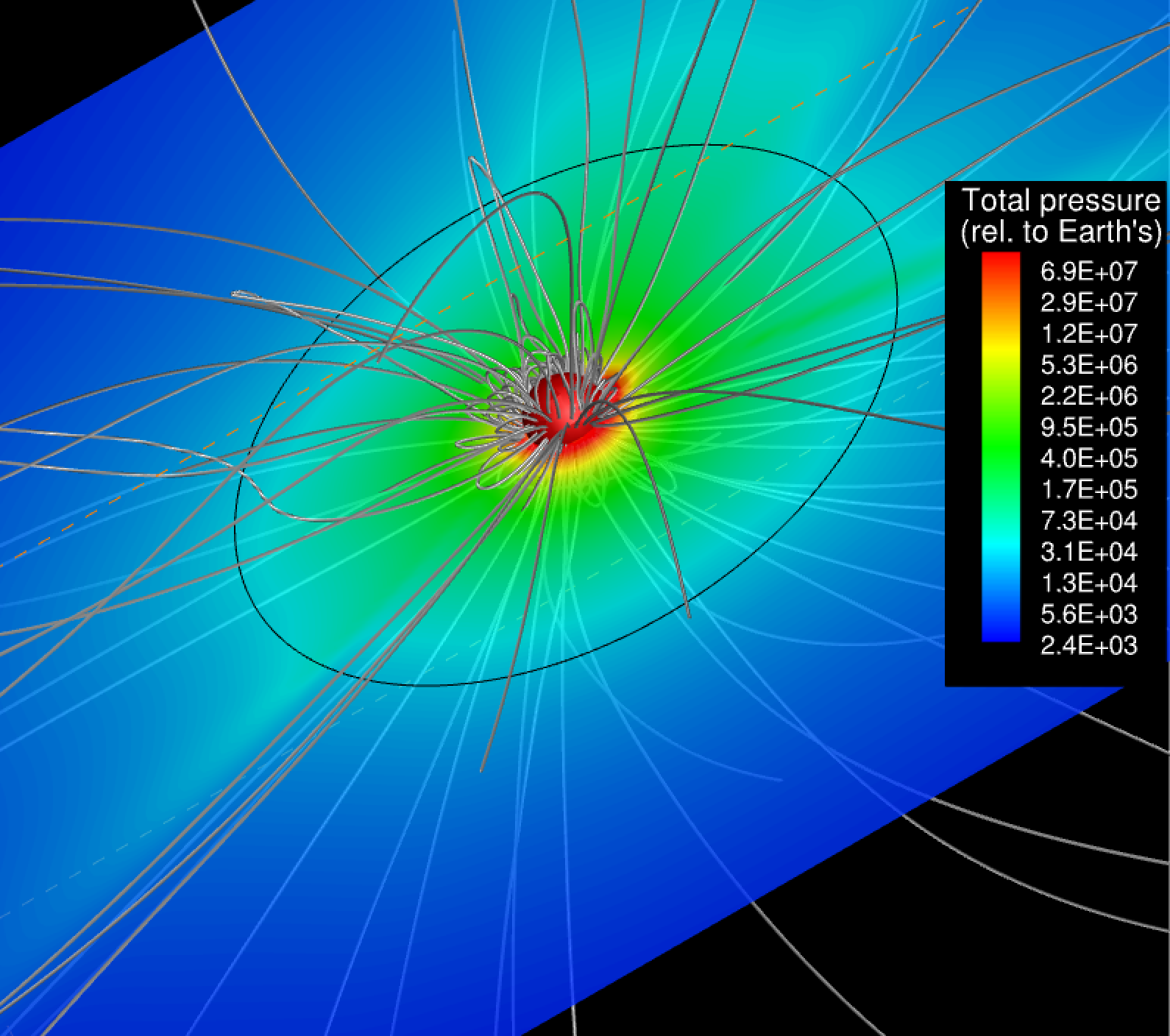}
\includegraphics[width=5.5cm]{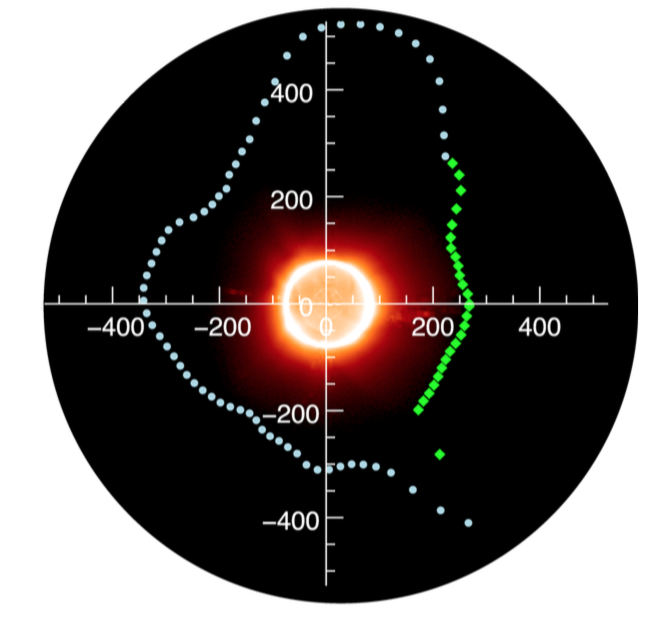}
\caption{Left: Stellar wind model of the planet-hosting star HD189733, computed using the observationally-derived ZDI magnetic map from \citet{2010MNRAS.406..409F}. The total wind pressure is shown in colour, in a slice through the equatorial plane, and is given in units relative to the solar wind pressure at the Earth's orbit. The orbital radius of HD\,189733b is represented by the black circumference. Right: The X-ray emission of the quiescent corona (i.e., it neglects the non-thermal emission of flares and associated to small-scale magnetic field components) is shown in the background. In the foreground, it is shown the  velocity of the stellar wind at the position of the orbit of HD\,189733b. Image from Llama et al. (2013, MNRAS, 436, 2179).}
\label{fig.av_wind}       
\end{figure}

One of the advantages of global wind models is that the numerical grid can extend out to large distances, allowing us to characterise the \hbindex{stellar wind} conditions around exoplanets \citep{2009ApJ...703.1734V,2010ApJ...720.1262V,2012MNRAS.423.3285V, 2015MNRAS.449.4117V,2011ApJ...733...67C,2011ApJ...738..166C,2013MNRAS.436.2179L,2016MNRAS.459.1907N,2017arXiv170303622V}. The characterisation of the stellar wind (i.e., the \hbindex{interplanetary medium}) is important to quantify the wind (magnetic and particles) effects on exoplanets, as we will discuss later. The global wind models can also incorporate more complex magnetic field topologies, including those derived from observations of \hbindex{stellar magnetism} (ZDI maps). The magnetic maps are imposed as boundary conditions at  the stellar wind base. The magnetic field lines are then extrapolated into the computational domain (e.g., corona, astrosphere), initially assuming the field is in its lowest energy state (i.e., a potential field). With the interaction of the stellar wind particles, the magnetic field becomes stressed. The self-consistent interaction between magnetic field lines and stellar wind particles are let to evolve, until a relaxed solution is found \citep[for more details, see e.g.,][]{2014MNRAS.438.1162V}. The left panel of Figure \ref{fig.av_wind} illustrates the solution of the stellar wind model of the planet-hosting star HD\,189733, computed using the observationally-derived ZDI magnetic map from  \citet{2010MNRAS.406..409F}. Colour-coded is the total wind pressure (thermal, magnetic and ram pressures) relative to the solar wind pressure at the Earth's orbit.  The right panel of Figure \ref{fig.av_wind} shows, in the background, the X-ray emission of the hot, quiescent corona of the star due to thermal free-free radiation \citep{2013MNRAS.436.2179L}. Coronal \hbindex{X-ray emission} comes mainly from flaring magnetic loops with different sizes. As the small-scale magnetic structure is not resolved in ZDI observations, the X-ray emission computed in \citet{2013MNRAS.436.2179L}  captures only the quiescent corona and, as such, provide a lower limit for the emission. Overlaid to the X-ray image in Figure \ref{fig.av_wind} is the velocity of the stellar wind at the position of the orbit of HD\,189733b (indicated in the left panel by the black circumference). 

%%%%%%%%%%%%%%%%%%%%%%%%%%%%%%%%%%%%%%%%%%
\section{Stellar wind effects on exoplanets}\label{sec.av_exoplanets}
The majority of exoplanets known nowadays orbit stars at considerably close distances. These close orbits are not represented in our solar system. The giant exoplanets at close-in orbits are also known as hot-Jupiters. The system presented in Figure \ref{fig.av_wind} and studied by \citet{2013MNRAS.436.2179L}, for instance, hosts a hot-Jupiter, namely HD\,189733b. The solid black line shown in the left panel of Figure \ref{fig.av_wind} represents its orbital radius, at about $\sim 8.7~R_\star$. From the values of the stellar wind pressure at the orbital position of this exoplanet, we see that the environment surrounding hot-Jupiters have considerably different physical conditions than those around the solar system planets. \hbindex{Stellar wind} densities, magnetic field intensities, temperatures, and pressures all decay with distance, albeit with different dependencies \citep[see also][]{2015MNRAS.449.4117V}. This means that the interactions between winds or stellar magnetic field lines are strongest for close-in planets. Despite the fact that stellar winds  are still accelerating and therefore  have usually low velocities at the position of close-in exoplanets, the orbital velocities of these planets are large due to the $1/\sqrt{r}$-dependence of Keplerian velocities. 
\hbindex{Close-in planets} can, therefore, have supersonic orbital velocities in the azimuthal direction \citep{2010ApJ...722L.168V}. This means that the relative velocity of the close-in planet through the wind of its host star can be as large as that of the solar wind impacting on the Earth's magnetosphere. The main difference between these two scenarios is the direction of the velocity vector, which has a large azimuthal (radial) component in the former (latter) case. 

Because of the large stellar wind densities at the position of \hbindex{close-in planets}, the relative motion of the planet through the wind of the star results in large ram pressures. This, for example, can give rise to the formation of bow shocks surrounding exoplanets \citep{2010ApJ...722L.168V, 2011MNRAS.411L..46V, 2013MNRAS.436.2179L,2013ApJ...764...19B}. In a recent numerical study, \citet{2015A&A...578A...6M} showed that other structures, such as cometary-type tails, and inspiraling accretion streams, can also appear as a result of the interactions between stellar winds and planetary magnetic fields/outflows \citep[see also][]{2010ApJ...721..923L,2010Natur.463.1054L,2013A&A...557A.124B,2016A&A...591A.121B}.

One consequence of the large pressure of the stellar wind environment around exoplanets is that it can constrain the sizes of planetary magnetospheres. The stand-off distance between the planetary surface and the magnetopause is set by pressure balance. At the planet--stellar wind interaction zone, pressure balance between the stellar wind (left-hand side) and planetary magnetosphere (right-hand side) can be written as
\begin{equation}\label{eq.equilibrium}
\rho \Delta u^2 +  p_{\rm th} + \frac{B_{r_{\rm orb}}^2}{8\pi}\simeq \frac{B_{{p},r_M}^2}{8\pi} ,
\end{equation}
where $\rho$, $p_{\rm th}$ and $B_{r_{\rm orb}}$ are the density, thermal pressure and magnetic field strength of the stellar wind at the position of the planetary orbit, $\Delta u$ is the relative velocity of the planet through the wind of the host-star and  $B_{{p},r_M}$ is the planetary magnetic field intensity at a distance $r_M$ from the planet centre. Eq.~(\ref{eq.equilibrium}) neglects the planetary thermal pressure component on the right side. Because of the exponential decay of planetary densities, at the height of a few planetary radii, thermal pressure is usually negligible compared to the planetary magnetic pressure. We further take the \hbindex{planetary magnetic field} to be dipolar, such that $B_{{p},r_M} = B_{p, {\rm eq}} (R_p/r_M)^3$, where $R_p$ is the planetary radius and $B_{p, {\rm eq}}$ its surface magnetic field at the equator (i.e., half the value of the intensity at the magnetic pole). For a planetary dipolar axis aligned with the rotation axis of the star, the magnetospheric size of the planet is given by
\begin{equation}\label{eq.r_M}
 \frac{r_M}{R_p} = \left[ \frac{B_{p, {\rm eq}}^2}{8 \pi (\rho \Delta u^2 +  p_{\rm th}) + B_{r_{\rm orb}}^2} \right]^{1/6}.
\end{equation}
Therefore, a large stellar wind pressure (ram, thermal and/or magnetic, shown in the denominator of Equation \ref{eq.r_M}) acts to reduce the size of \hbindex{planetary magnetospheres} for a given planetary magnetisation \citep[see also, e.g., ][]{2004ApJ...602L..53I,2006JGRA..111.6203Z,2008MNRAS.389.1233L,2009A&A...505..339L,2009ApJ...703.1734V,2010ApJ...720.1262V,2012MNRAS.423.3285V,2011MNRAS.412..351V,2011JGRA..116.1217S,2012ApJ...744...70K,2013ApJ...765L..25B}.  This can have important effects on the habitability of exoplanets, including terrestrial type planets orbiting inside the habitable zones of their host stars \citep{
2005AsBio...5..587G,2009Icar..199..526G, 2007AsBio...7..167K, 2007AsBio...7..185L,2011MNRAS.412..351V,2011AN....332.1055V, 2010Icar..210..539Z,2013A&A...557A..67V,2014A&A...570A..99S,2016A&A...596A.111R}. The magnetosphere acts to deflect the stellar wind particles, preventing its direct interaction with planetary atmospheres and, therefore, atmospheric erosion.  If the \hbindex{magnetosphere} is too small, part of the atmosphere of the planet becomes exposed to the interaction with stellar wind particles. \citet{2007AsBio...7..185L} suggested that, for a magnetosphere to protect the atmosphere of planets, planets are required to have magnetospheric sizes of $\gtrsim 2~R_p$. 

\subsection{Effects on potentially habitable planets around M dwarf stars}
Due to a combination of a \hbindex{habitable zone} that is closer to the star and the technologies currently adopted in exoplanet searches, \hbindex{M dwarf (dM) stars} have been the prime targets for detecting terrestrial planets in the potentially life-bearing region around the star.  dM stars, however, can have significantly high magnetic fields, in particular, when they are young, fast rotating and, thus, active \citep{2008MNRAS.390..545D,2008MNRAS.390..567M,2010MNRAS.407.2269M, 2009ApJ...692..538R}. For example, the late dM star WX\,UMa, have large-scale magnetic fields of up to {4}~kG \citep{2010MNRAS.407.2269M}, significantly larger than the large-scale solar magnetism of several G \citep{2016MNRAS.459.1533V}. With time, the magnetic field intensity of dM stars decay \citep{2014MNRAS.441.2361V}. However, dM stars are known to remain active over long timescales \citep[on the order of Gyr,][]{2008AJ....135..785W}. This implies that the environment surrounding exoplanets that orbit dM stars should remain highly magnetised for long periods of time.  

\citet{2013A&A...557A..67V} studied the effects of the high magnetisation of dM stars in setting the sizes of the magnetospheres of (hypothetical) Earth-like planets orbiting in their habitable zones. Using a sample of 15 dM stars with measured surface magnetic fields and assuming planets to have a similar terrestrial magnetisation, they showed that these hypothetical Earths would have magnetospheres that extend as low as $1~R_p$ ($1.7~R_p$) and at most up to $6~R_p$ ($11.7~R_p$), for planets orbiting in the inner (outer) edge of the habitable zone. The magnetospheric size of Proxima b (orbiting around a M5.5V star, see Table \ref{table.proxima_centauri}) has been estimated to extend up to about $1.3$ to $7~R_p$ \citep{2016A&A...596A.111R}.  For comparison, the Earth's magnetosphere is located at  $r_M\simeq 10$ -- $15~R_\oplus$ \citep{1992AREPS..20..289B}, showing that Earth-like planets with similar terrestrial magnetisation orbiting active dM stars present {\it smaller} \hbindex{magnetospheric sizes} than that of the Earth. If exoplanets lack a protective magnetic shield, this potentially implies that these planets can lose a significant fraction of their atmospheres \citep{2013ApJ...770...23Z}. 

\subsection{Effects on potentially habitable planets orbiting young stars}
Sun-like stars are observed to emit larger X-ray and extreme ultra-violet (EUV) fluxes at a younger age \citep[e.g.][]{2003ApJ...594..561G,2005ApJ...622..680R,2010ApJ...714..384R}. The large stellar irradiation can heat the outer atmospheres of exoplanets that become more extended and more susceptible to evaporation \citep[e.g.][]{2003ApJ...598L.121L,2004A&A...419L..13B,2011A&A...532A...6S}. Planets can receive an enhanced flux of energetic photons if they are orbiting at close distances to their stars and/or if they are orbiting around young and more active stars.

In addition to larger X-ray and EUV fluxes, young and more active stars also harbour more intense magnetic fields and winds with larger $\dot{M}$. Relation (\ref{eq.wood}), between $\dot{M}$ and $F_x$, has important consequences, for example, for the evolution of the young solar System. Extrapolations using Eq.~(\ref{eq.wood}) suggest that the 700~Myr-Sun would have had $\dot{M}$ that is $\sim 100$ times larger than the current solar mass-loss rate $\dot{M}_\odot = 2 \times 10^{14}~\msano$ \citep{2005ApJ...628L.143W}. This could explain the loss of the Martian atmosphere as due to erosion caused by the stronger wind of the young Sun \citep{2004LRSP....1....2W}. 

Through modelling of the wind of the solar twin $\kappa$ Ceti, \citet{2016ApJ...820L..15D} computed the magnetospheric size of the \hbindex{young Earth}. $\kappa$ Ceti is believed to be a good representation of our Sun at an age of about $400$ -- $600$~Myr, roughly the time that life started emerging on our planet. To calculate the magnetospheric size of the (hypothetical) Earth-twin, we make use of Equation (\ref{eq.r_M}). For a similar present-day magnetisation of the Earth, \citet{2016ApJ...820L..15D} estimated the size of the magnetosphere of the \hbindex{young Earth} to be about $5~R_p$. Contrary to Mars, which does not host a significant magnetic field and therefore no appreciable magnetosphere, the relatively large size of the early Earth's magnetosphere  may  have been the reason that prevented the volatile losses from Earth and created conditions to support life.

%%%%%%%%%%%%%%%%%%%%%%%%%%%%%%%%%%%%%%%%%
\section{Final remarks}
In this Chapter, we presented one aspect of how stars can affect their surrounding planets. We concentrated on the effects that stellar winds might have on exoplanets. We only presented case studies in which the exoplanet hosts a protective magnetic field, which prevents a direct interaction between the stellar wind and the planetary outer atmosphere. Non-magnetised exoplanets are believed to undergo significant atmospheric erosion in short timescales. However, exoplanetary magnetic fields have not yet been directly observed and so far have only been elusively probed \citep[e.g.][]{2008ApJ...676..628S,2010ApJ...722L.168V,2014Sci...346..981K}. Due to the diversity of stellar and exoplanetary properties, of the architectures of exoplanetary systems and the complex nature of stellar wind--planet interaction,  progress in this field is likely to come through different angles. This is another example of how exoplanetary studies are becoming more multi-disciplinary, where important physical insights can only be gained from efforts arising from different areas of Astrophysics, such as stellar,  solar,  and planetary Physics, and astrobiology.

%\begin{svgraybox}
%If you want to emphasize complete paragraphs of texts we recommend to use the newly defined Springer class option and environment \verb|svgraybox|. This will produce a 15 percent screened box `behind' your text.
%\end{svgraybox}

%\hbindex{indexed term}.

%\section{Cross-References}
% indicate the titles of these chapters, but please not more than 10 of them.
% 
%\begin{itemize}
%\item{Other worlds in ancient greek philosophy}
%\item{Exoplanet detection in the 21st century}
%\item{Fundamental limits to knowledge on exoplanets}
%\end{itemize}

%\begin{acknowledgement}
%Optionally, include a short acknowledgment.  Else, out-comment this section. Not more than a few lines please.
%\end{acknowledgement}

\bibliographystyle{spbasicHBexo}  %for bibtex

%%%%%%%%%%%%%%%%%%%%%%%%%%%%%

\end{document}